\documentclass[aps,pra,epsfigure,twocolumn,longbibliography]{revtex4-1}
\usepackage{dcolumn}    
\usepackage{bm} 
\usepackage{graphicx}
\usepackage{amssymb}
\usepackage{amsmath}    
\usepackage{latexsym}
\usepackage{amsfonts}   
\usepackage{array}      
\usepackage{epsfig}
\usepackage{txfonts}
\usepackage{color}
\usepackage[colorlinks=true,linkcolor=blue,urlcolor=blue,citecolor=blue,pdfusetitle]{hyperref}
\usepackage{hyperref}

\newcommand{\ket}[1]{\left\vert#1\right\rangle}

\newcommand{\bassano}[1]{{\color{black} #1}}

\newcommand{\mean}[1]{\langle #1 \rangle}

\begin{document}
\title{Nonequilibrium quantum bounds to Landauer's principle: Tightness and effectiveness}
\author{Steve Campbell,$^{1,2}$ Giacomo Guarnieri,$^{2,1,3}$  Mauro Paternostro,$^{4,5}$ and Bassano Vacchini$^{2,1}$}
\affiliation{
$^1$Istituto Nazionale di Fisica Nucleare, Sezione di Milano, via Celoria 16, 20133 Milan, Italy\\
$^2$Dipartimento di Fisica, Universit{\`a} degli Studi di Milano, via Celoria 16, 20133 Milan, Italy\\
$^3$Department of Optics, Palack\'{y} University, 17. listopadu 1192/12, 771 46 Olomouc, Czech Republic\\
\mbox{$^4$Centre for Theoretical Atomic, Molecular and Optical Physics, Queen's University Belfast, Belfast BT7 1NN, United Kingdom}\\
\mbox{$^5$Laboratoire Kastler Brossel, ENS-PSL Research University, 24 rue Lhomond, F-75005 Paris, France}}
\begin{abstract}
We assess two different nonequilibrium quantum Landauer bounds: the traditional approach based on the change in entropy, referred to as the `entropic bound', and one based on the details of the dynamical map, referred to as the `thermodynamic bound'. By first restricting to a simple exactly solvable model of a single two level system coupled to a finite dimensional thermal environment and by exploiting an excitation preserving interaction, we establish the dominant role played by the population terms in dictating the tightness of these bounds with respect to the dissipated heat, and clearly establish that coherences only affect the entropic bound. Furthermore, we show that sharp boundaries between the relative performance of the two quantities emerge, and find that there are clear instances where both approaches return a bound weaker than Clausius' statement of the second law, rendering them ineffective. Finally we show that our results extend to generic interaction terms.
\end{abstract}
\date{\today}
\maketitle

\section{Introduction}
Landauer's principle provides us with the fundamental conclusion that information is physical and its erasure is necessarily accompanied by a minimum thermodynamic cost, the dissipated heat~\cite{Landauer1961,Bennett1982a}. It is now increasingly accepted that the assessment of genuinely quantum systems and elementary quantum processes necessitates the re-examination of familiar thermodynamic quantities such as work and heat~\cite{JohnReview}. With this in mind, it is quite remarkable that Landauer's principle extends beyond its original classical paradigm and equally applies when the state of a joint system-environment configuration is quantum~\cite{AndersPRE}. 

A clear understanding of how a quantum system dissipates heat is intrinsically important both from a fundamental and practical standpoint. Indeed, such disordered forms of energy are a potential source of inefficiency in emerging quantum technologies. Thus recently, several studies have explored lower bounds on the dissipated heat in a variety of systems, including the experimental tests of Landauer's principle~\cite{BerutNature,JunPRL,PetersonPRSA,RoccoArXiv}, examining the validity of Landauer's bound for a fully quantum setting~\cite{AndersPRE}, its behavior in open quantum systems~\cite{PezzuttoNJP,RuariPRL}, schemes to minimize the dissipated heat~\cite{OmarNJP}, and a rigorous tightening of Landauer's bound~\cite{ReebWolfNJP}. While Landauer's principle is rooted in the use of information-theoretic entropies~\cite{Landauer1961, Bennett1982a,ReebWolfNJP}, recent studies have shown that other approaches that do not necessarily invoke any information theoretic tools but rather rely on the dynamics of the system, can be used to derive a ``nonequilibrium thermodynamic" lower bound on the dissipated heat~\cite{GooldPRL,GiacomoArXiv}. \bassano{The relevance of this approach further relies on the fact that a microscopic analysis of the erasing procedure allows to take into account effects related to non-Markovianity or initial correlations \cite{Breuer2016a}, which have often shown to lead to counterintuitive phenomena \cite{Campbell2012a,Man2012a}.

 Despite their quite different origin, both entropic and thermodynamic bounds are valid, though possibly far from being tight, for a generic system-environment interaction. It is therefore of interest to ascertain the relative performance of the bounds, while also exploring their dependence on the choice of initial system state and environmental temperature. Indeed the dependence of the erasing procedure and the related entropy variation and dissipated heat on the initial system state was one of the basic concerns already considered in the seminal paper by Landauer \cite{Landauer1961}. A better understanding of the relationship of the two bounds, in particular in their dependence on the initial state, further provides hints on the interplay between logical and thermodynamic irreversibility. Indeed the latter issue is all the more relevant in the quantum framework due to the different role of measurement in quantum mechanics.  
 
 It is exactly in this direction that this work progresses. We examine the relative performance of Landauer's entropic bound with the bound derived in Ref.~\cite{GooldPRL}. To this aim we consider a system consisting of a single qubit coupled to a finite-dimensional thermal environment.} Initially assuming an excitation-preserving interaction, we show that the details of the initial system state are crucial in dictating the tightness of the different bounds. Remarkably, we find that in the parameter space of the initial system, sharp boundaries emerge, highlighting a cross-over between the bounds. Interestingly, the presence or absence of coherences is shown to play a diminished role, only entering into the details of the entropic bound and being completely absent from the dissipated heat as well as the thermodynamic bound. Furthermore, it is shown that the same qualitative behavior extends beyond the excitation-preserving interactions to more generic models.

The remainder of this paper is organised as follows. In Sec.~\ref{LandauerBounds} we define the lower bounds on the dissipated heat that will be the focus of this work. Sec.~\ref{XXsection} we exhaustively study the performance of these bounds in a simple excitation preserving model. In Sec.~\ref{OtherModels} we show the results persist for generic interaction terms. Finally, Sec.~\ref{conclusions} we present our conclusions and a short discussion.

\section{Landauer-type Bounds}
\label{LandauerBounds}
Consider a situation in which the total Hamiltonian of a system in contact with an environment is time-independent, such that no work is done. The heat dissipated by the system into its environment can be expressed as
\begin{equation}
\label{eq:DissHeat}
\mean{Q} = \text{Tr}\left[ \mathcal{H}_E \left( \varrho_E(t) - \varrho_E(0) \right)  \right],
\end{equation}
where $\varrho_E$ is the density operator, and $\mathcal{H}_E$ the Hamiltonian of the environment. Using an information theoretic framework, Landauer established that this quantity can be bounded from below by examining the corresponding change in entropy
\begin{equation}
\label{eq:Landauer}
\beta \langle Q \rangle \geq \Delta S = S(\varrho_S(0)) - S(\varrho_S(t)),
\end{equation}
where $\beta$ is the inverse temperature, $S(\cdot)$ is the von Neumann entropy and $\varrho_S$ is the density operator of the system. For brevity we refer to Refs.~\cite{Landauer1961,ReebWolfNJP} for a more detailed discussion. This result, which has recently been tightened when quantum systems are explicitly considered by Reeb and Wolf~\cite{ReebWolfNJP}, is remarkable as it was one of the first instances to explicitly demonstrate the physical nature of information. In the following we will refer to Eq.~\eqref{eq:Landauer} as the ``entropic bound".

The growing interest in exploring the thermodynamics of quantum systems~\cite{JohnReview} has led to a closer examination of Landauer's principle and the dissipated heat~\cite{ReebWolfNJP,GooldPRL,RuariPRL,PezzuttoNJP,GiacomoArXiv,OmarNJP,JunPRL,KoskiPRL,RoccoArXiv,PetersonPRSA,DahlstenPRL,CiampiniNPJQI,LutzScience}. Recently, a different approach to bounding $\beta\mean{Q}$ was proposed in Ref.~\cite{GooldPRL}. Starting from the unitary dynamics of the total system-environment state and employing a heat fluctuation relation, the dissipated heat can be bounded by a quantity that is related to the dynamical map governing the evolution of the system. Explicitly it was shown that~\cite{GooldPRL}
\begin{equation}
\label{eq:MJ}
\beta \langle Q \rangle \geq \mathcal{B} = -\ln\left(\text{Tr}\left[ \sum_i K_i^\dagger \varrho_S(0) K_i \right] \right)
\end{equation} 
with $K_i$ the Kraus operators of the map acting on the system, which depend on the environment initial state (assumed here to be in Gibbs form) as well as the system-environment interaction Hamiltonian. Again, for brevity we refer to Ref.~\cite{GooldPRL} for a detailed derivation. We will refer to this as the ``thermodynamic bound". 

Clearly, the approaches used to derive the bounds are fundamentally different in nature and this leads us to explore their respective relevance in the dependence on choice of initial states and features of the dynamics. To answer this question we will examine these quantities in simple exactly solvable systems, showing that the answer reveals remarkably subtle features of the two approaches. 

\begin{figure}[t]
{\bf (a)}\\
\includegraphics[width=0.8\columnwidth]{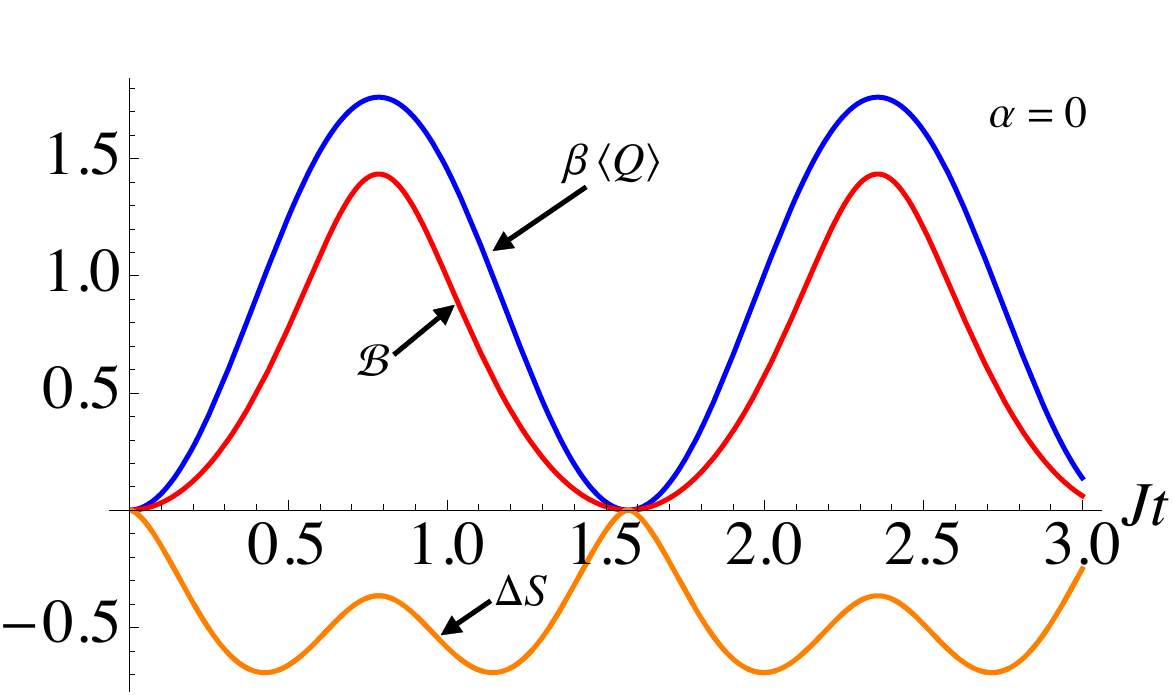}\\
{\bf (b)}\\
\includegraphics[width=0.8\columnwidth]{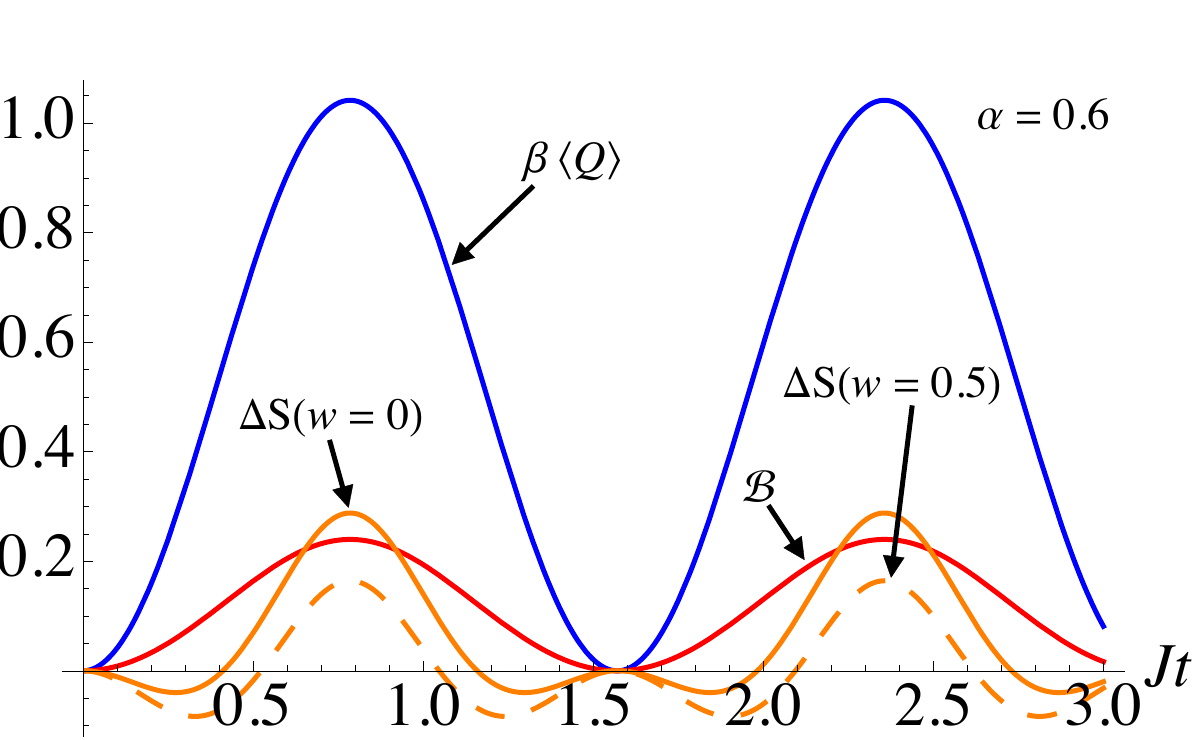}
\caption{Dissipated heat $\beta \langle Q \rangle$ (blue), thermodynamic bound $\mathcal{B}$ (red) and Landauer bound $\Delta S$ (orange). In both panels we set the inverse temperature of the environment qubit $\beta=1$. {\bf (a)} Pure excited initial system state, i.e. $\alpha=0$. {\bf (b)} Mixed initial system state with $\alpha=0.6$. The solid orange curve is for a fully dephased initial state $w=0$ while the dashed orange curve is for $w=0.5$. We remark that the thermodynamic bound and the dissipated heat have no dependence on $w$.}
\label{fig1}
\end{figure}
\section{Description of the system-environment coupling: XX-Interaction}
\label{XXsection}
Our model consists of two coupled qubits, which we label system, $S$, and environment, $E$, with free Hamiltonians $\mathcal{H}_{S(E)}\!=\!\sigma^z$.
As a first characterization of the model the two spins are coupled via an $XX$ interaction
\begin{equation}
\mathcal{H} = J (\sigma^x_S \otimes \sigma^x_E + \sigma^y_S \otimes \sigma^y_E  ),
\end{equation}
with $\sigma_i$ the usual Pauli matrices, where the coupling is measured in energy units set by the free evolution. The environment qubit is initially in a thermal state $\varrho_E(0)=e^{-\beta \mathcal{H}_{E}}/\mathcal{Z}$ with $\mathcal{Z}$ the associated partition function. As noted in Ref.~\cite{GooldPRL}, the case of a single spin environment is already sufficient to capture the salient features of the quantities at hand, while still allowing for a fully analytical treatment. Furthermore, it serves as a benchmark for larger interacting systems which will be considered elsewhere. We take the initial state of the system to be
\begin{equation}
\label{eq:SYS}
\varrho_S(0) = \left(
\begin{array}{cc}
 1-\alpha ^2 & \delta \\
\delta & \alpha ^2 \\
\end{array}
\right),~~~\text{with}~~~\delta=w \left( \alpha  \sqrt{1-\alpha ^2} \right),
\end{equation}
where $0\leq w\leq 1$ and $0\leq \alpha ^2\leq 1$, so that $0\leq \delta\leq 1/2$.
Note that we are assuming the ordered basis $\left\{\ket{1},\ket{0}\right\}$. The simplicity of the interaction allows us to readily determine Eqs.~\eqref{eq:DissHeat}-\eqref{eq:MJ}, however given their somewhat involved form we omit explicitly reporting them here. In Fig.~\ref{fig1} we examine the effect taking different initial states for the system has on the relative performance of the bounds. \bassano{This simple analysis already allows us to infer that both the dissipated heat, Eq.~\eqref{eq:DissHeat}, and the thermodynamic bound, Eq.~\eqref{eq:MJ}, are independent of the value of $w$, i.e. the presence of coherence in the initial state has no bearing on these quantities.} For an initially pure state, as in panel {\bf (a)}, we find $\Delta S\!\leq\!0$ and its dynamical behavior bears little affinity to that of the dissipated heat, while the thermodynamic bound closely mimics $\beta\mean{Q}$. However, the fact that $\mathcal{B}$ is a tighter bound when the initial state is pure is simply an artefact of the special restriction such an initial state puts on $\Delta S$: as the state is pure the entropy can only increase and therefore Eq.~\eqref{eq:Landauer} is always negative. Considering an initially mixed system state, panel {\bf (b)}, we see the situation becomes much more subtle. While the thermodynamic bound still closely tracks the functional behavior of the dissipated heat it is not as tight as in the pure excited state case. Furthermore, since the initial state is mixed the entropy can now both increase and decrease due to the interaction. We now see that if $w=0$ (solid orange curve) corresponding to a fully dephased initial system state, then during the dynamics the entropic bound can be tighter than the thermodynamic bound, while for a partially dephased state (dashed orange) the converse can hold true.

\begin{figure}[t]
{\bf (a)}\\
\includegraphics[width=0.8\columnwidth]{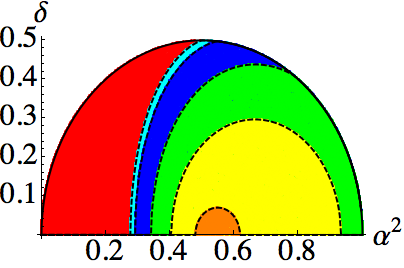}\\
 {\bf (b)}\\
 \includegraphics[width=0.8\columnwidth]{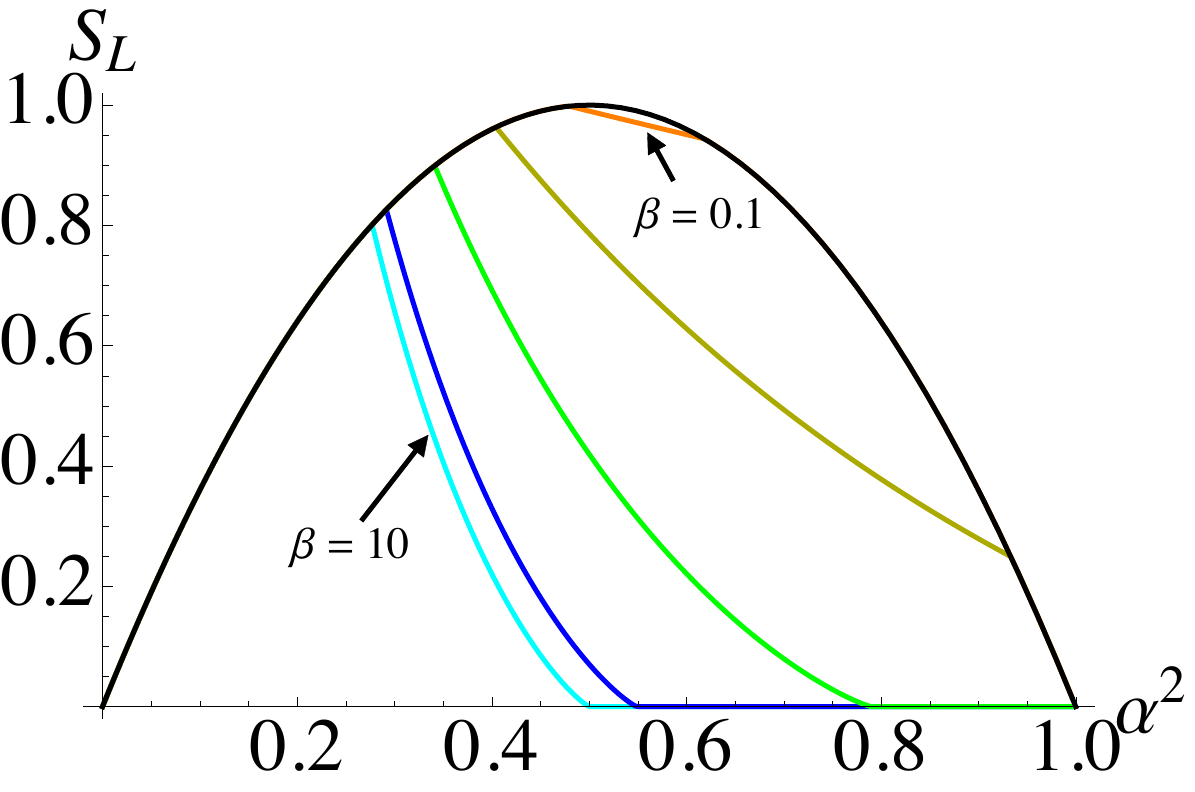}
\caption{{\bf (a)} The leftmost red region contains states for which $\mathcal{B}_\text{max}\!>\!\Delta S_\text{max}$ for $\beta=10$, while for all other states $\Delta S_\text{max}\!>\!\mathcal{B}_\text{max}$. As the temperature of the environment is increased we find the range of states for which $\Delta S_\text{max}\!>\!\mathcal{B}_\text{max}$ shrinks and is delineated by everything within the colored regions (left to right): $\beta=10$ [cyan], 2 [blue], 1 [green], 0.5 [yellow], and 0.1 [orange]. {\bf (b)} Linear entropy of the states lying along the boundaries given by the dashed lines in panel {\bf (a)}.}
\label{fig2}
\end{figure}

\subsection{Quantitative comparison of the bounds}
Comparing the relative performance of the two bounds is delicately dependent on the particular details of the initial state of the overall system. Since the coupling is excitation preserving, the total system has a very well defined period which we can exploit to make our analysis more quantitative. The question still arises: how do we unambiguously define which bound is tighter? While clearly there are the trivial instances at which $\beta \langle Q \rangle\!=\!0$ and both quantities are tight, we see that dynamically they can exhibit crossovers. Thus, even though for arbitrary times the interplay between the bounds is complex, it is arguably most interesting to determine which is tighter when the dissipated heat is maximized, $\beta\mean{Q}_\text{max}$. Due to the periodicity this simply means determining the value of the bounds at time $t=\pi/(4J)$. Concise analytical expressions putting into evidence the different role of coherences and populations can be obtained using the standard Bloch representation for the statistical operator $\rho_{S} = \frac{1}{2} ( \openone + \bm{\sigma}\cdot \mathbf{v} )$, with $\mathbf{v}=(v_x, v_y ,v_{z} )$. Taking into account the invariance of the dynamics with respect to the choice of polar angle in the Bloch sphere, corresponding to reality of~Eq.~\eqref{eq:SYS}, we take $\mathbf{v}=\left( 2 \delta , 0,1-2 \alpha^{2} \right)$, thus obtaining
\begin{equation}
\label{eq:MJMax}
  \mathcal{B}_{\max} = - \ln [ 1-v_{z} \tanh ( \beta ) ],
\end{equation}
and
\begin{equation}
\label{eq:LandauerMax}
  \Delta S_{\max} = \Delta S_{\beta} -\Delta S_{v} .
\end{equation}
where
\begin{equation*}
\begin{aligned}
&\Delta S_{\beta} = & \beta \tanh \beta  + \ln \sqrt{1- \tanh^{2} \beta}  \\
&\Delta S_{v} = & \ln \sqrt{1- | v |^{2}} + | v | \text{tanh}^{-1}| v | .
\end{aligned}
\end{equation*}
We remark that at this instant in time the interaction plays the role of a swap operation between the state of the system and the environment.
Note that an important difference between the bounds already appears in these expressions. As noted previously, we explicitly see that the thermodynamic bound does not depend on the coherences, and is determined by the interplay between the environmental temperature and the initial population of the system's ground state. On the other hand, we see the entropic bound is affected by coherences, but interestingly it is the sum of two contributions. The first is independent on the state of the system and provides a fixed offset ranging from 0 to a plateau at its maximum value $\ln 2$ for decreasing environmental temperature. The second term depends on the initial system state only through the modulus of the Bloch vector.

Using Eqs.~\eqref{eq:MJMax} and \eqref{eq:LandauerMax} we explore the role that the temperature of the environment and the initial state of the system has on the relative performance of both bounds in Fig.~\ref{fig2}. In panel {\bf (a)} we randomly generate millions of initial states for $\varrho_S$ and determine which is closer to $\beta\mean{Q}_\text{max}$. Setting $\beta=10$ corresponds to a cold environment such that it is essentially initialised in its ground state. The leftmost red region shows the states for which $\mathcal{B}_\text{max}>\Delta S_\text{max}$, while in the (lighter) cyan region and for all further states to the right we find the converse. The dashed lines show the boundary states, that can be found by solving the transcendental equation $\mathcal{B}_\text{max} \!=\! \Delta S_\text{max}$. While both quantities have a clear temperature dependence, for the entropic bound the decrease in the system's entropy calls for a growth of the environmental entropy, favoured by the purity of its initial state. In particular, the temperature dependence of $\Delta S_\text{max}$ is typically much weaker than compared with $\mathcal{B}_\text{max}$, and furthermore this contribution is independent from the system's initial state. It follows then that the region in which the entropic bound out performs the thermodynamic bound shrinks and progressively tends towards a point corresponding to the maximally mixed initial system state. At this point it is important to stress a caveat regarding Fig.~\ref{fig2} {\bf (a)}: as evidenced previously, within the parameter space one or both bounds can be negative, even when the dissipated heat is positive. Therefore, there are regions in which one bound outperforms the other, however ultimately both fail to provide any meaningful information regarding the dissipated heat. We will return to this point more explicitly in the proceeding section.

Given the invariance of $\mathcal{B}$ to the presence of coherences, which clearly appears from Eq.~\eqref{eq:MJMax}, we can conclude that the main parameter delineating the regions is the initial populations, parameterized by $\alpha$. The dominate role that the populations play in dictating the performance of the bounds is shown in Fig.~\ref{fig2} {\bf (b)}, where we re-parameterize panel {\bf (a)} to show the linearized entropy $S_L = 2 \left(1 - \text{Tr} \left[ \varrho_S^2 \right] \right)$ against the ground state population $\alpha^2$ for the boundary states, i.e. those states lying along the dashed lines. States lying on the $x-$axis correspond to pure states while those along the outer boundary are the maximally mixed for a given value of $\alpha$. Clearly, for predominantly excited ($\alpha^2\lesssim0.5$) the thermodynamic bound is always tighter regardless of the temperature of the environment.

\subsection{Tightness of the bounds}
\begin{figure}[t]
{\bf (a)}\\
\includegraphics[width=0.85\columnwidth]{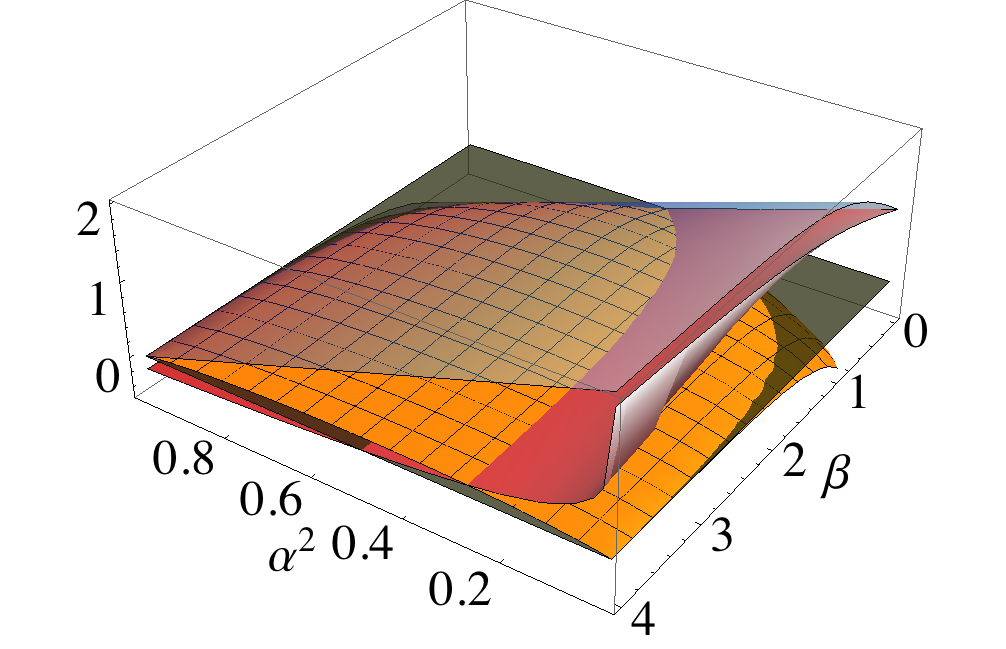}\\
{\bf (b)}\\
\includegraphics[width=0.75\columnwidth]{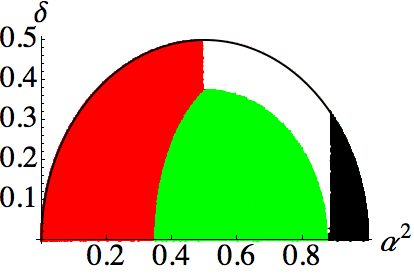}\\
{\bf (c)}\\
\includegraphics[width=0.75\columnwidth]{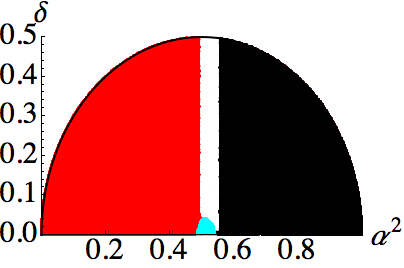}\\
\caption{{\bf (a)} Analysis of the tightness of the bounds. We plot $\mean{Q}_\text{max}$ [topmost], $\mathcal{B}_\text{max}/\beta$ [red], and $\Delta S_\text{max}/\beta$ [meshed, orange] for $w=0$. The flat plane at zero is for reference. {\bf (b)} and {\bf (c)} We re-examine which bound is tighter with the additional constraints that the dissipated heat and both bounds are positive. The black regions correspond to where $\beta \mean{Q}<0$ and therefore is not a relevant implementation of Landauer's principle. The white region is when $\beta \mean{Q}>0$ but $\mathcal{B}_\text{max}<0$ and $\Delta S_\text{max}<0$. The leftmost red region represents states such that $\mathcal{B}_\text{max}>\Delta S_\text{max}$, while $\mathcal{B}_\text{max}<\Delta S_\text{max}$ for $\beta=1$ [central green region in panel {\bf (b)}] and $\beta=0.1$ [central cyan region in panel {\bf (c)}].}
\label{fig3}
\end{figure}

While the previous section highlights which bound serves as a better estimate for $\beta\mean{Q}$, an immediate question arises: how close do either of these quantities get to the dissipated heat? From Fig.~\ref{fig1} we see instances where the thermodynamic bound is close to the actual dissipated heat, while it appears the entropic bound is always quite loose. In Fig.~\ref{fig3} {\bf (a)} we rescale the quantities by $\beta$ for clarity and examine their tightness for the extremal value of $w=0$ corresponding to fully dephased states which maximizes the entropic bound. The top-most transparent plane is $\mean{Q}$, the red and meshed orange planes correspond to $\mathcal{B}_\text{max}/\beta$ and $\Delta S_\text{max}/\beta$, respectively, and finally the dark flat plane is at zero. We see that for $\alpha^2=0$ which corresponds to a pure excited system state, $\mathcal{B}$ serves as a reasonable lower bound on $\beta\mean{Q}$, and in fact for $\beta \to 0$ becomes tight, although we remark this corresponds to an infinite temperature environment and therefore the actual dissipated heat $\beta\mean{Q}\to 0$. Changing $\alpha$ the discrepancy grows between $\mathcal{B}_\text{max}/\beta$ and $\mean{Q}_\text{max}$, and we find for $\alpha^2>0.5$ the thermodynamic bound is always negative. By fixing $w=0$ we see that for cold environments $\Delta S_\text{max}$ only provides a better bound when both quantities are quite far from the true value of the dissipated heat. For any $w\neq0$ the entropic bound performs progressively worse and for $w=1$ is always negative. 

This behavior highlights a further point mentioned previously: with the exception of a small region of the parameter space, namely $\alpha^2 \geq \tfrac{1}{2}\left[ 1+ \tanh(\beta) \right]$, the dissipated heat is positive $\beta \mean{Q} \geq 0$. While evidently $\mathcal{B}_\text{max}<0$ if $\alpha^2>0.5$ and $\Delta S_\text{max}$ can be negative for a wide range of parameter choices. From the Clausius statement of the second law, it is immediate to conclude that
\begin{equation}
\label{clausius}
\beta \mean{Q} \geq 0~~~\text{for}~~~\alpha^2 \leq \tfrac{1}{2}\left[ 1+ \tanh(\beta) \right].
\end{equation}
Thus there can be clear situations in which both bounds fail to capture any features of the dissipated heat. It is therefore of relevance to compare $\Delta S_\text{max}$ and $\mathcal{B}_\text{max}$ in the parameter region in which the dissipated heat is positive, and in which they provide a more informative statement than Clausius' law, which we do in Fig.~\ref{fig3} {\bf (b)} and {\bf (c)} for $\beta=1$ and 0.1, respectively. The black regions show the states for which $\beta \mean{Q}<0$ and therefore are not a relevant implementation of Landauer's principle. Conversely, the white region shows the states in which {\it both} quantities are negative despite $\beta \mean{Q}>0$. Hence, in these regions neither bound is any more informative than Clausius' law, Eq.~\eqref{clausius}. Notice that a lower environmental temperature reduces the size of these regions significantly, and we remark for $\beta\gtrsim10$ we find $\beta \mean{Q}>0$ (since the environment is essentially in its ground state) and one bound is always positive. Thus, while the results of Fig.~\ref{fig2} {\bf (a)} explicitly show which bound is tighter, it excludes the instances when either Landauer's principle does not hold or when both bounds are weaker than Clausius' law. While the vertical sharp boundaries in Fig.~\ref{fig3} {\bf (b)} and {\bf (c)} are a consequence of the independence of $\beta \mean{Q}$ and $\mathcal{B}$ with respect to $w$, the emergence of the sharp cross-overs between all the regions is nevertheless remarkable.

\section{description of the system-environment coupling: Other Interaction Models}
\label{OtherModels}
In order to understand the general features in the interplay between the bounds we generalize our results to other interaction models showing that the previously obtained qualitative features indeed persist for arbitrary interactions. In order to do so we must modify our strategy to compare the bounds, as arbitrary interactions do not exhibit such a clean periodic behavior, thus there is no definite point during the dynamics where a simple comparison can be made. Therefore we will consider the average value of the bounds, taken over the coupling, as one would naturally do in a operational approach to comply with a fluctuating interaction strength
\begin{equation}
\label{averaged}
\overline{\mathcal{A}} = \int_0^{J_\text{max}} dJ \, p(J)\,\mathcal{A}_{J}.
\end{equation}
The bounds are evaluated at a time much longer than the free evolution time, so as to ensure that they have reached an asymptotic value. Due to the significantly more involved nature of the quantities, when evaluating the bounds, $\overline{\mathcal{B}}$ and $\overline{\Delta S}$, these averages are performed numerically by taking a large sample of random values for $J$ from the interval $(0,J_\text{max})$ and taking the statistical mean. Considering an Ising interaction between the qubits 
\begin{figure}[t]
{\bf (a)}\\
\includegraphics[width=0.85\columnwidth]{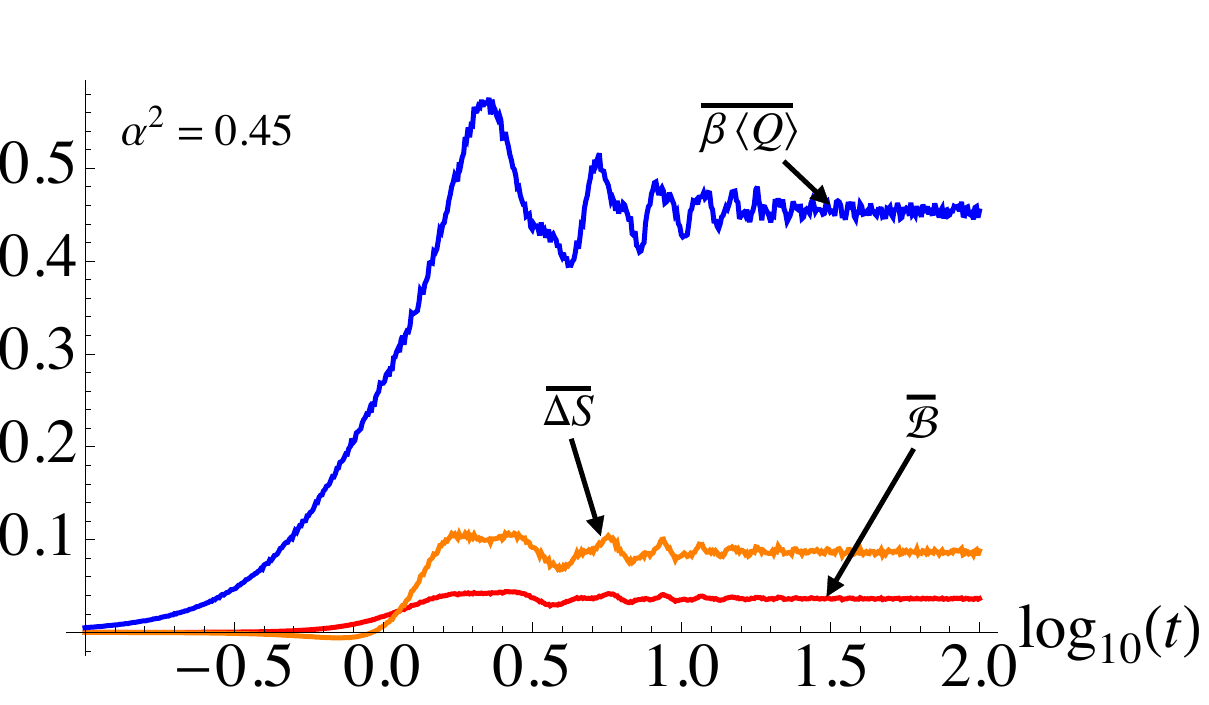}\\
{\bf (b)}\\
\includegraphics[width=0.85\columnwidth]{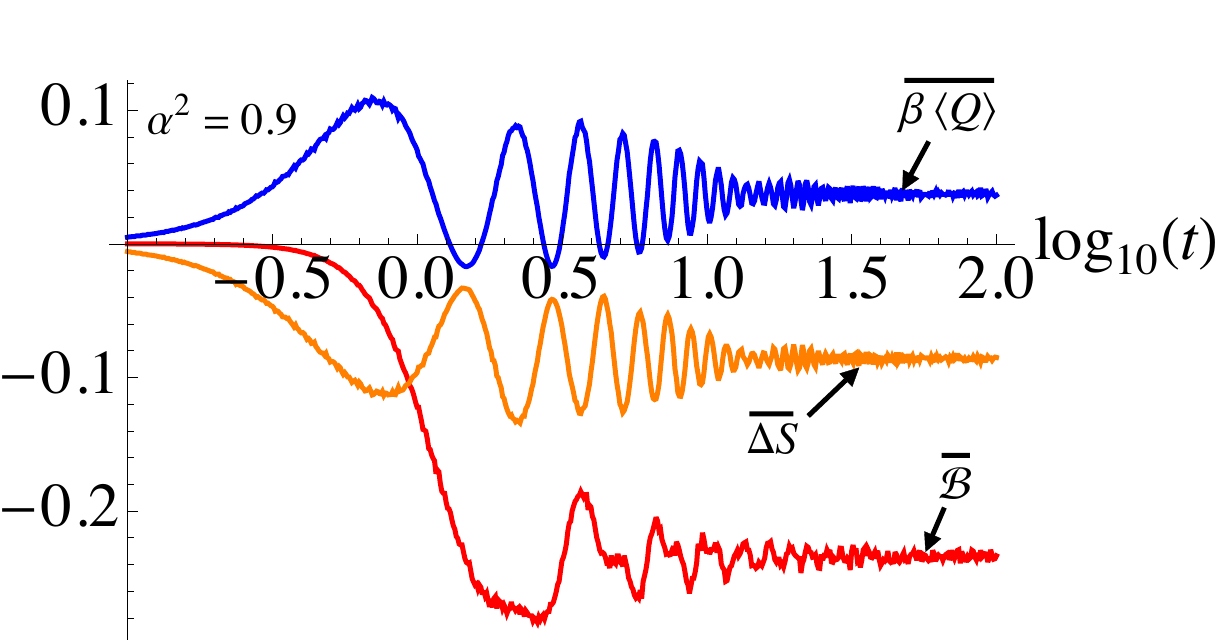}\\
{\bf (c)}\\
\includegraphics[width=0.85\columnwidth]{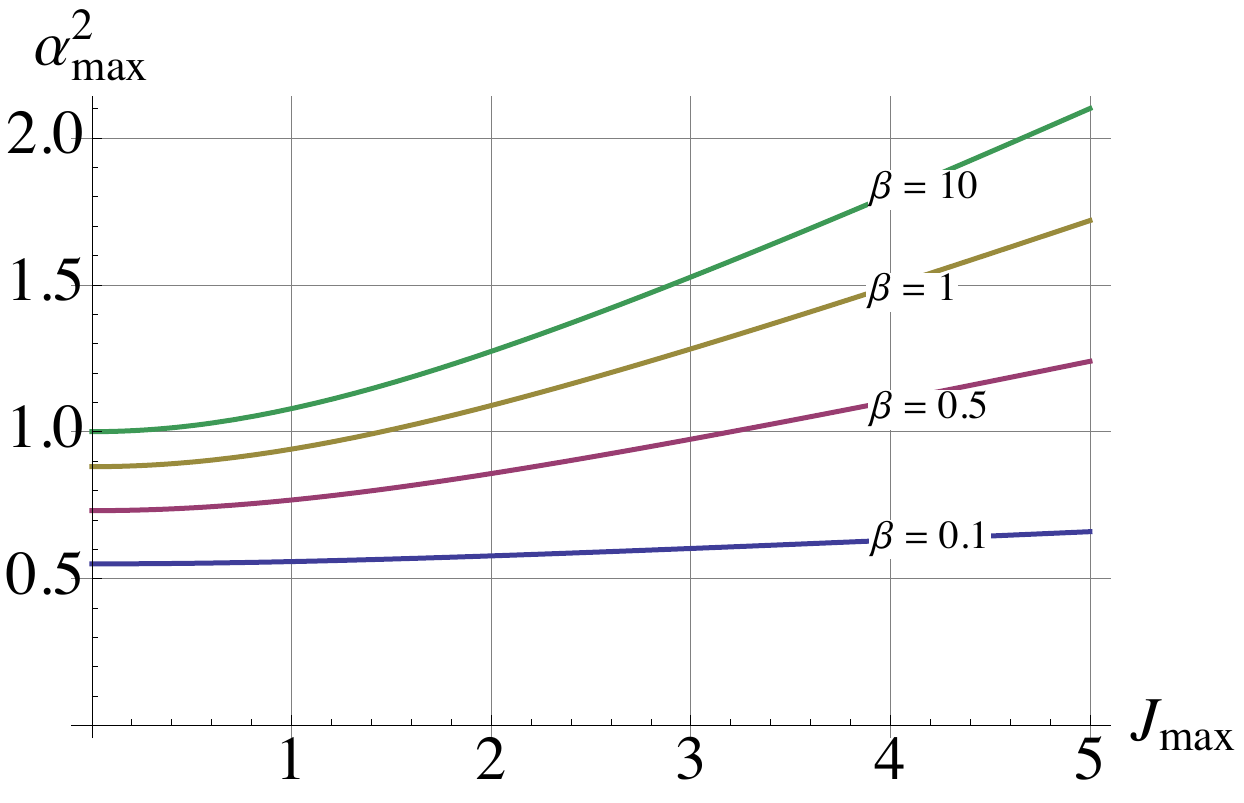}
\caption{Fixing $\beta=1$, $J_\text{max}=1$, and ensuring the coherence term of the initial state $\delta=0.1$ we examine the averaged values of the quantities for {\bf (a)} $\alpha^2=0.45$ and {\bf (b)} $\alpha^2=0.9$. {\bf (c)} The maximum allowed value of initial system ground state population, $\alpha^2_\text{max}$, such that the averaged dissipated heat, $\overline{\beta\mean{Q}}\geq0$.}
\label{fig4}
\end{figure}
\begin{equation}
\label{eq:Ising}
\mathcal{H} = J \sigma^x_S \otimes \sigma^x_E,
\end{equation}
in Fig.~\ref{fig4} {\bf (a)} and {\bf (b)} we show that already $t=10^2$ when $J=1$ is sufficient to ensure good convergence. Furthermore, we immediately see several qualitative features carry over to the new interaction model. In particular, we again find that the thermodynamic bound becomes negative when $\alpha^2 >0.5$ and furthermore find that the entropic bound is the only one sensitive to coherences. In the same way as for the $XX$ model, we can establish a bound on the value of the ground state initial population, $\alpha^2$, such that the dissipated heat is positive and therefore represents a meaningful instance of Landauer's principle. For the dissipated heat, due to the comparative simplicity of Eq.~\eqref{eq:DissHeat} we can evaluate Eq.~\eqref{averaged} analytically in the long time limit and we find that Clausius' law holds when
\begin{equation}
\label{eq:alphabound}
\alpha^2 \leq \frac{1}{2}\left[ 1 +\tanh(\beta)\left( \frac{J_\text{max}}{\arctan(J_\text{max}/2)}-1 \right)\right].
\end{equation}
This result immediately puts into evidence the non-excitation preserving nature of the Ising interaction. Examining Fig.~\ref{fig4} {\bf (c)} we see inline with intuition that if the environment is initially cold then the dissipated heat is always positive. However, for hotter environments we see that taking a larger value of $J_\text{max}$ is sufficient to ensure the dissipated heat is always positive, due to the fact that the interaction term is now injecting significant amounts of energy into the total system. 

\begin{figure}[t]
{\bf (a)}\\
\includegraphics[width=0.85\columnwidth]{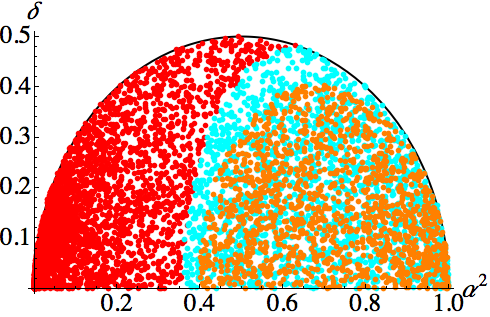}\\
{\bf (b)}\\
\includegraphics[width=0.85\columnwidth]{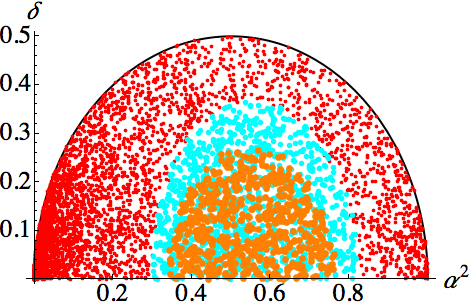}
\caption{{\bf (a)} ${J_\text{max}}=1$ and {\bf (b)} ${J_\text{max}}=10$. Similarly to Fig.~\ref{fig2}, in both panels the red regions $\overline{\mathcal{B}}\!>\! \overline{\Delta S}$ for $\beta=10$, while for all other states $ \overline{\Delta S} \!>\! \overline{\mathcal{B}}$. Hotter environments lead to smaller regions in which $\Delta S_\text{max}\!>\!\mathcal{B}_\text{max}$ as indicated by everything within the colored regions: $\beta=10$ [lighter cyan], 1 [darker orange]. For each point we set $t=10^3$ and average over $500$ values for $J$.}
\label{fig5}
\end{figure}

Following from the previous analysis we determine $\overline{\mathcal{B}}$ and $\overline{\Delta S}$ for various initial states of the system and for different initial temperatures for the environment and examine which bound is tighter. Sampling $J\!\in\!(0, {J_\text{max}})$, Fig.~\ref{fig5} {\bf (a)} shows that qualitatively the same behavior is exhibited as found for the $XX$ model when the coupling ${J_\text{max}}=1$. Again, the thermodynamic bound outperforms the entropic bound for initially highly excited states. As the temperature of the environment is increased, the range of states such that $\overline{\Delta S}> \overline{\mathcal{B}}$ shrinks. Furthermore, we recall these figures must be caveated inline with Eq.~\eqref{eq:alphabound} where $\overline{\beta\mean{Q}}<0$, and more significantly where both bounds are negative despite the dissipated heat being positive, cf. Fig.~\ref{fig4} {\bf (b)}. Hence, we conclude that virtually all of the qualitative features exhaustively explored for the $XX$ model, where an analytical treatment was possible, extend to other interactions models when the interaction term is of the same order of magnitude as the systems natural energy, i.e. for our purposes when $J_\text{max}\approx1$. We find some important differences arising for stronger couplings due to the non-excitation preserving nature of a generic interaction term.

More specifically, the non-excitation preserving nature of the interaction leads to the results shown in Fig.~\ref{fig5} {\bf (b)}. For large $J_\text{max}\gtrsim 5$, we know the dissipated heat is positive for a wide range of $\beta$, cf Fig.~\ref{fig4} {\bf (c)}.  Setting $J_\text{max}=10$, we see the division between which of the bounds is tighter changes significantly. While there is still the appearance of sharp crossvers between the performance of the two quantities, now they have become more symmetrical around the central region of the plot. Note that we still have that $\overline{\mathcal{B}}<0$ for $\alpha^2>0.5$ while the behavior of $\overline{\Delta S}$ is more involved. We remark however that, for the Ising case, both bounds are typically quite far from the average value of the dissipated heat $\overline{\beta\mean{Q}}$ and therefore act as only a loose bound. 

Although we have focused on the Ising model, the preceding analysis can be performed for a generic two-body interaction of the form $\mathcal{H} = \sum_{k=x,y,z} J_k (\sigma^k_S \otimes \sigma^k_E)$, with random couplings $J_k$ involving all the Pauli operators. In this case, the computational resources are significantly increased due to the more involved interaction term. Regardless, we obtain analogous results to those shown in Fig.~\ref{fig2} and ~\ref{fig5}. In particular, sharp convex subsets of the Bloch sphere marking a crossover between the performance of the two bounds, which are qualitatively the same as those obtained for the $XX-$ and Ising interactions, thus indicating that our results are robust against any choice of two-body interaction in one-dimensional systems.

\section{Conclusions}
\label{conclusions}
We have compared and contrasted different formulations of nonequilibrium quantum Landauer bounds. We have shown the delicate dependence of the `entropically' defined and `thermodynamically' defined bounds to the initial state of the system and environmental temperatures. Remarkably, the thermodynamic formulation shares several features with the dissipated heat, in particular its independence to the presence of initial state coherences, a feature not shared by the entropic approach. By examining the relative performance of the quantities we find sharp boundaries exist in the parameter space, and more interestingly there are instances where both are negative despite the dissipated heat being positive. In these situations the bounds are weaker than the standard Clausius' statement of the second law. The features explored in this work were exhaustively shown for an excitation preserving interaction, however the qualitative behavior was confirmed to persist for generic interaction models.

Of course given that the bounds are derived from disparate formalisms, it is not surprising that they should perform differently. However, as our results highlight, there are interesting subtleties when one explicitly considers how they perform for a generic initial system state and different environmental conditions. It is quite remarkable the seemingly small role quantum coherences play in both instances and it is interesting to consider if this extends to entangled systems undergoing Landauer-like erasure. It is important to note that the analysis could be performed using the sharpened entropic bound derived by Reeb and Wolf~\cite{ReebWolfNJP}, however, the qualitative features shown here remain since this bound adds a correction to Eq.~\eqref{eq:Landauer} rather than significantly changing it. Furthermore, the more involved form of this bound would have largely ruled out an insightful analytical treatment.

\acknowledgements
We acknowledge support from the EU Collaborative projects QuProCS (grant agreement 641277) and TherMiQ (grant agreement 618074), the UniMi H2020 Transition Grant, and the Julian Schwinger Foundation (grant number JSF-14-7-0000). G.G. acknowledges the support of the Czech Science Foundation (GACR) (grant no. GB14-36681G). MP is supported by the DfE-SFI Investigator Programme (grant 15/IA/2864) and the Royal Society Newton Mobility Grant NI160057. This work was partially supported by the COST Action MP1209.

\bibliography{Landauer_bib_v7}
\end{document}